\documentstyle[12pt]{article}
\begin{document}
\title {GROUND STATES FOR THREE-LAYER HYBRID SUPERCONDUCTOR}
\author {by\\MA{\L}GORZATA SZTYREN
\thanks{Department of Mathematics and Information Science,
Warsaw University of Technology, Pl. Politechniki 1, PL-00-661 Warsaw,
E--mail: emes@mech.pw.edu.pl}}
\date \today
\maketitle

\newcommand {\eqn}[1]{\begin{equation}#1\end{equation}}
\newcommand {\eqna}[1]{\begin{eqnarray}#1\end{eqnarray}}
\newcommand{\equln}[2]{\begin{equation} {#1} \label{#2} \end{equation}}
\newcommand{\eqnl}[2]{\begin{equation} {#1} \label{#2} \end{equation}}
\newcommand {\la}{\longrightarrow}

\begin{abstract}
   A superconducting hybrid structure composed of three layers is
considered. The 2D layers interact mutually by higher grade interlayer
couplings. We determine the possible superconducting modes. Those solutions
enable to discuss the conditions for the onset and enhancement
of 3D superconductivity in such a structure.
\end{abstract}
\section{Introduction}
    We are concerned with a system of potentially superconducting
2D layers interacting with one another by possibly distant interlayer
couplings. The general aim of our considerations is the study of
the nanoelectronic properties of such structures, including the
dependence of their electronic properties on the structural and
mechanical parameters.\\
    We shall make use of the higher grade hybrid model (HM) of 
supercoductors formulated in \cite{Sztyren:2003} and developed in 
\cite{ Rogula:2004, Sztyren:2006, Rogula+Sztyren:2006,Sztyren:2006a,
Sztyren:2007}
In the present paper we confine ourselves to three-layer structures,
with the aim to determine the complete set of solutions.\\
  According to \cite{Sztyren:2003}, in the framework of HM
the layered superconductor is considered
as a one-dimensional chain with "atoms" (being here identical atomic
planes described by 2D Ginsburg-Landau theory with parameters $\alpha_0$ and
$\beta$) and with Josephson's bonds (called J-links) between them.
The grade \(K\), expressed by an arbitrary (but specified for any
particular case) integer, defines the admitted range of Josephson's
interaction in terms of interplanar gaps.

   We denote by \(\psi_n\) the order
parameter associated to the layer indexed by the number \(n\). Its
complex conjugate (c.c.) is denoted by \(\bar{\psi}_n\).
In the considered variant of HM (identical
atomic planes), besides the GL parameters characterising isolated
planes
there are two classes of long-range coupling
constants: Josephson parameters \(\gamma_q\) and proximity efect 
parameters \(\zeta_q\). In the present paper we intend to examine
the influence
of these long-range coupling, under assumption that there is no
magnetic field and no currents.\\

    Following \cite{Sztyren:2003} we shall now briefly present the
equations for plane-uniform states of our hybrid model
in the absence of magnetic field. The order parameter is then
independent of the in-plane variables and the net supercurrents
vanish. 
We shall confine our attention to the grade
\(K=2\).
    For \(K=2\) the condition of vanishing Josephson current
is equivalent to
\begin{equation}
\gamma_1(\bar\psi_n\psi_{n+1}-c.c.)
 +\gamma_2(\bar\psi_n\psi_{n+2}
 +\bar\psi_{n-1}\psi_{n+1}-c.c.)=0,
\label{vani}
\end{equation}
and the field equations take the form
\begin{equation}
\tilde{\alpha}_n\psi_n +\beta|\psi_n|^2\psi_n-\frac{1}{2}[\gamma_1
(\psi_{n+1}+\psi_{n-1})+\gamma_2(\psi_{n+2}+\psi_{n-2})]=0,
\label{far}
\end{equation}
where instead of $\alpha_0$ we have introduced
\equln{\tilde{\alpha}_n=\alpha_0+\frac{1}{2}\sum_q\zeta_q
,}{c311}
with the summation runing over the planes which are J-linked
to the plane number $n$. The plane index $n$ belongs to a given set
$P$.
\section{Exact solutions for N=3}
   Let us consider the simplest non-trivial instance of higher grade
system composed of a finite number of layers:
the case \(K=2\) (hence interactions of nearest and next nearest
neighbours
only, characterized by parameters \(\zeta_1,\, \gamma_1,\,
\zeta_2,\,\gamma_2\)), the number of layers equal 3.
We shall index the planes with the integers from the set 
\(P=\{-1, 0, 1\}\).
The interlayer gaps will be then numbered by \(-1/2\) and \(1/2\).
The energy functional takes the form
\begin{equation}
\begin{array}{ll}
{\cal{E}}=
&\tilde{\alpha}_1|\psi_1|^2+\tilde{\alpha}_{-1}|\psi_{-1}|^2
+\tilde{\alpha}_0|\psi_0|^2-\frac{\gamma_1}{2}[(\bar{\psi}_{-1}+
\bar{\psi}_{1})\psi_0+\bar{\psi}_{0}(\psi_{-1}+\psi_{1})]\\\\
&-\frac{\gamma_2}{2}(\bar{\psi}_{-1}\psi_{1})+\bar{\psi}_{1}\psi_{-1})
+\frac{\beta}{2}(|\psi_1|^4+|\psi_0|^4+|\psi_{-1}|^4)\\
\end{array}
\label{E1}
\end{equation}
and the
system of equations for the order parameters \(\psi_{-1},\ \psi_0\)
and \(\psi_1\) reads
\equln{(\tilde{\alpha}_1+\beta|\psi_1|^2)\psi_1-\frac{1}{2}
 [\gamma_1\psi_0+\gamma_2\psi_{-1}]=0,}
{c3_1}
\equln{(\tilde{\alpha}_0+\beta|\psi_0|^2)\psi_0-\frac{1}{2}
 \gamma_1(\psi_1+\psi_{-1})=0,}
{c3_2}
\equln{(\tilde{\alpha}_{-1}+\beta|\psi_{-1}|^2)\psi_{-1}-\frac{1}{2}
 [\gamma_1\psi_0+\gamma_2\psi_1]=0.}
{c3_3}
Due to the finitenes of the system of layers under consideration,
the conditions (\ref{vani}) for vanishing Josephson's currents
\(J_{-\frac{1}{2}}\) and \(J_{\frac{1}{2}}\)
are automatically satisfied. For the present system they take the form
\equln{\gamma_1\bar{\psi}_{-1}\psi_0+ \gamma_2\bar{\psi}_{-1}\psi_1- c.c.=0
,}{j12}
\equln{\gamma_1\bar{\psi}_0\psi_1+ \gamma_2\bar{\psi}_{-1}\psi_1- c.c.=0
.}{j_12}

   Note that, according to (\ref{c311}),
\(\tilde{\alpha}_1\) and \(\tilde{\alpha}_{-1}\) are equal
to one another; hence we shall use the symbol \(\tilde{\alpha}_1\)
for both the cases. We have
\equln{\tilde{\alpha}_1=\alpha_0+\frac{1}{2}(\zeta_1+\zeta_2)
              =\tilde{\alpha}_0-\frac12\delta
,}{alf1}
where
\equln{\delta=\zeta_1-\zeta_2.
}{delta}
For the central layer we have
\equln{\tilde{\alpha}_0=\alpha_0+\zeta_1
.}{alf0}
   The trivial solution \(\psi_j=0,\ \  j\in P \) describes
the normal state. We are interested in discussing the stability of
the normal solution, and in finding stable nontrivial
solutions to the system (\ref{c3_1}-\ref{c3_3}) describing
the superconducting states. 
\section{Properties}
  The eqns. (\ref{c3_1}-\ref{c3_3}) have a number of specific
properties which facilitate complete solution of the system.
To compactify the phrasing, we shall use
the following terminology.

   Definition 1. A solution to system (\ref{c3_1}-\ref{c3_3})
is called {\em T-invariant}
iff it is gauge-equivalent to a real solution \(\psi_j\):
\equln{\bar{\psi}_j=\psi_j.}{tinv}

   Definition 2. A solution to system (\ref{c3_1}-\ref{c3_3})
is called {\em TP-invariant}
    iff it can be gauge-transformed to the form
\equln{\bar{\psi}_{-j}=\psi_j.}{tpinv}

   It is also convenient to introduce the symbols for real and 
imaginary parts of the order parameters:
\equln{\psi_j=a_j+ib_j,\ \ \ \ j\in P
.}{ab}

   Property 1. Any solution to eqns. (\ref{c3_1}-\ref{c3_3})
is gauge-equivalent to a solution which satisfies the following
conditions
\eqnl{\psi_0=a_0
}{gau}
and
\equln{b_1+b_{-1}=0
.}{b11}
Moreover, any such solution fulfills
\equln{b_1[\gamma_1a_0+\gamma_2(a_1+a_{-1})]=0
.}{jj}

   Property 2. Any real solution satisfies  
\equln{(a_1-a_{-1})[\alpha^*_1+\beta(a_{1}^2+a_1a_{-1}+a^2_{-1})]=0
.}{c3_13}
where
\equln{\alpha^*_1=\tilde{\alpha}_0+\frac{1}{2}(\gamma_2-\delta)
.}{b111}

   Property 3. For any solution \(\psi_0=0\) iff
\(\psi_1=\psi_{-1}\).

   Property 4. Any solution which is not T-invariant,
is TP-invariant, and vice versa.

\section{Classification of solutions}
   Due to the Property 1 we can confine further consideration
to the nontrivial solutions satisfying (\ref{gau}) and (\ref{b11}).
From the remaining properties it follows that this set of solutions
to the system (\ref{j12}-\ref{c3_3}) is
partitoned into the following four disjoint classes:\\\\
\indent Class (A): \(b_1=0,\ \ a_1=-a_{-1}\,{\neq}\,0,\ \ a_0=0\),\\ 
\indent Class (B): \(b_1\,{\neq}\,0,\ \ a_1=a_{-1},\ \ a_0\,{\neq}\,0\),\\
\indent Class (C1): \(b_1=0,\ \ a_1=a_{-1},\ \ a_0\,{\neq}\,0\),\\
\indent Class (C2): \(b_1=0,\ \ |a_1|\,{\neq}\,|a_{-1}|,\ \ a_0\,{\neq}\,0 \).\\

The classes (A) and (C2) are T-invariant, the class (B) is 
TP-invariant, and the class (C1) is T- and TP-invariant. The classes
(B) and (C2) are 3D, the class (A) is 1D and the class (C1) is 2D.
   In the respective classes, the field equations
(\ref{j12}-\ref{c3_3}) reduce to simplified forms. Only equations of
the class (B) contain imaginary parts.\\

   In the class (A) the situation is particularly simple. We have
%
\equln{a_1^2=-\frac{\alpha^*_1}{\beta}
,}{a1}
so that (taking into account the positiveness of \(\beta\))
the class is non-empty iff
\equln{2\tilde{\alpha}_0+\gamma_2-\delta<0
.}{lin}
In the remaining classes the situation is more complicated.\\

Class (B): The equations take the form
\equln{a^2_0=-\frac{1}{\beta}(\tilde{\alpha_0}+
             \frac{\gamma_1^2}{2\gamma_2})
,}{a0}
\equln{a_1=a_{-1}=-\frac{\gamma_1}{2\gamma_2}a_0
,}{a11c}
\equln{b^2_1=|\psi_1|^2-a^2_1=b^2_{-1}
,}{b12}
where
\equln{|\psi_1|^2=|\psi_{-1}|^2=-\frac{\alpha^*_1}{\beta}
}{ampli}
and \(\alpha^*_1\) is given by the eqn. (\ref{b111}).
The necessary and sufficient condition for the existence of 
solutions in this class is the conjunction of (\ref{lin}) and
the following two inequalities
\equln{\tilde{\alpha}_0+\frac{\gamma^2_1}{2\gamma_2}<0
,}{hiper}
\equln{(\gamma^2_1-4\gamma^2_2)\tilde{\alpha}_0
       -2\gamma^2_2(\gamma_2-\delta)+
       \frac {\gamma^4_1}{2\gamma_2}>0
.}{super}

Class (C1):
 The system of equations for \(a_0\) and \(a_1\) can be written as
\equln{a_0=\frac 2{\gamma_1}(\alpha^*_1-\gamma_2+\beta a_1^2)a_1
,}{c1}
\equln{a_1=\frac 1{\gamma_1}(\tilde{\alpha}_0+\beta a_0^2)a_0
,}{c2}
where $\alpha^*_1$ is given by the eqn. (\ref{b111}). 
Let us multiply
by sides the equations (\ref{c1}) and (\ref{c2}). The result is
\equln{(\alpha_1^*-\gamma_2+\beta a_1^2)(\tilde{\alpha}_0+\beta a_0^2)
=\frac 12\gamma_1^2
.}{multi}
Now we make the ansatz
\equln{\tilde{\alpha}_0+\beta a_0^2=x
,}{anx}
and
\equln{2(\alpha_1^*-\gamma_2+\beta a_1^2)=\gamma_1^2\frac{1}{x}
.}{an1x}
We are looking for non-zero solutions of the equations (\ref{c1})
and (\ref{c2}), hence we can divide the equations by sides.
We obtain
\equln{\beta a_0^2=x-\tilde{\alpha}_0>0
}{betaa0}
and
\equln{\beta a_1^2=\gamma_1^2\frac{1}{x}-\alpha_1^*-\gamma_2>0
,}{betaa1}
where both the inequalities formulate the necessary conditions
for the existence of solutions in the class (C1).
Further
\equln{\gamma_1^2(\frac{a_1}{a_0})^2=x^2
,}{x2}
and, taking into account (\ref{b111}),
\equln{\frac{x^2}{\gamma_1^2}=
\frac{\frac{\gamma_1^2}{2x}-\tilde{\alpha}_0+
\frac 12(\delta+\gamma_2)}{x-\tilde{\alpha}_0}
.}{x22}
Geometrically, the set of solutions from class (C1) are determined
by the points of intersection of two curves:
\equln{y=\frac{2}{\gamma_1^2}x^3
,}{y1}
and
\equln{y=\frac{\gamma_1^2-2(\tilde{\alpha}_0
-\delta-\gamma_2)x}{x-\tilde{\alpha}_0}
.}{y2}
Let us note, that $x=\tilde{\alpha}_0$ corresponds to the zero-solution,
describing the transition to the normal state. We have then the
equation for $\tilde{\alpha}_0$
\equln{2(\tilde{\alpha}_0-\delta-\gamma_2)\tilde{\alpha}_0
-\gamma_1^2=0
,}{Ns}
which always has two real roots. We shall denote them by
\equln{\alpha_{01}=\frac 14(\delta+\gamma_2-\sqrt{(\delta+\gamma_2)^2+
8\gamma_1^2})
,}{al1}
and
\equln{\alpha_{02}=\frac 14(\delta+\gamma_2+\sqrt{(\delta+\gamma_2)^2+
8\gamma_1^2})
,}{al2}
For a given material, the number of solutions in the class (C1) 
depends on the temperature parameter $\tilde{\alpha}_0$
in the following manner.\\
\indent 1) When $\tilde{\alpha}_0<\alpha_{01}$, there are two solutions\\
\indent 2) When $\alpha_{01}<\tilde{\alpha}_0<\frac 12(\delta+\gamma_2)$ or
 $\frac 12(\delta+\gamma_2)<\tilde{\alpha}_0<\alpha_{02}$,\\
\indent \indent
there exists one solution\\
\indent 3) For $\alpha_{02}<\tilde{\alpha}_0$ there are no solutions.\\\\

Class (C2):
   The solutions of this class fulfill the condition
\equln{a_{1}^2+a_1a_{-1}+a^2_{-1}=-\frac{\alpha^*_1}{\beta}
.}{b2}
By combining it with real parts of eqns. (\ref{c3_1})-(\ref{c3_3})
we obtain
\equln{a_1=\frac 12(a_++a_-),\ \ \ a_{-1}=\frac 12(a_+-a_-)
,}{apm}
where the variables \(a_+,\ a_-\) satisfy the equations
%
\equln{a^2_-=-4\frac{\alpha^*_1}{\beta}-3a^2_+
,}{b21}
\equln{a_0=-\frac 2{\gamma_1}(\alpha^*_1+\frac{\gamma_2}2+\beta a_+^2)a_+
,}{ca1}
\equln{a_+=\frac 2{\gamma_1}(\tilde{\alpha}_0+\beta a_0^2)a_0
.}{ca2}
Because of (\ref{b2}), the inequality (\ref{lin}) is the necessary
condition for the existence of solutions in the class (C2).
To simplify the discussion of the sufficient conditions let us first 
introduce some additional symbols:
\equln{\alpha_1=\frac{2\alpha^*_1+\gamma_2}{\gamma_1},\ \ \ \ 
    \lambda=\alpha_1^4,\ \ \ 
     \kappa=\frac14\frac{\tilde\alpha_0}{\gamma_1}\alpha_1
}{alfa12}
and once more change
variables. We shall express the quotient and the product of $a_+$ and
$a_0$ by $\xi$ and $\eta$ defined as follows.
\equln{\xi=\frac{a_+}{a_0}\alpha_1,\ \ \ \eta=
         \frac{2\beta}{\gamma_1}\frac {1}{\alpha_1^2}a_+a_0
.}{ksieta}
Then the equations (\ref{ca1}) and (\ref{ca2}) will be transformed
into
\equln{\xi^2-4\kappa\xi-\lambda\eta=0
}{etaxi1}
and
\equln{\eta\xi^2+\xi+1=0
.}{etaxi2}
Eliminating $\eta$ one obtains the following equation for $\xi$
\equln{\xi^4-4k\xi^3+l(\xi+1)=0
,}{ksi}
which can be transformed into the form
\equln{w^4=(w-p)^2-q
,}{parabole}
by introducing the variable:
\equln{w=\sqrt{6}\kappa(\xi-\kappa)
}{defw}
and the coefficients:
\equln{p=\frac{\sqrt{6}}{36}\frac{\lambda-8\kappa^3}{2\kappa^3},\ \ \\
       q=p^2+\frac{-3\kappa^4+\kappa\lambda+\lambda}{36\kappa^4}
.}{defpq}
The request of tangency of curves
\equln{\eta=\frac{1}{\lambda}(\xi^2-4\kappa\xi)
}{eta1}
and
\equln{\eta=-\frac{\xi+1}{\xi^2}
}{eta2}
implies the condition
\equln{2\xi^4-4\kappa\xi^3-\lambda(\xi+2)=0
.}{stycz}
Combining (\ref{stycz}) with (\ref{ksi}) one can express $\kappa$
and $\lambda$ by $\xi$:
\equln{\lambda=\frac{\xi^4}{2\xi+3}
}{ell}
and
\equln{\kappa=\frac{\xi(3\xi+4}{2\xi+3}
.}{kap}
Now we can solve (\ref{kap}) with respect to $\xi$:
\equln{\xi_{1,2}=\frac 13(\kappa-2\mp\sqrt{(k+2)^2+2})
}{ksi12}
and then, apropriately substituting $\xi_1$ or $\xi_2$, calculate
$\lambda_{1,2}$ as well as $\eta_{1,2}$. The curves
(\ref{eta1}) and (\ref{eta2}) are tangent to each other at points
$(\xi_1,\eta_1)$ and $(\xi_2,\eta_2)$. For $\lambda > \lambda_1$
there are no points of intersection of curves 
(\ref{eta1}) and (\ref{eta2}), hence no superconducting solutions to
equations (\ref{ca1}) and (\ref{ca2}). For $\lambda = \lambda_1$
there are two solutions.
Finally, for $\lambda > \lambda_1$ there are four solutions.\\
\section{Onset of superconductivity}
   Let us first discuss the stability of the normal state.
Due to Properties the second variation of energy can be represented
in the form
\begin{equation}
\begin{array}{ll}
\delta^2{\cal{E}}=
&(\tilde{\alpha}_1+2\beta|\psi_1|^2)[(\delta a_1)^2+(\delta b_1)^2]+\\\\
&(\tilde{\alpha}_1+2\beta|\psi_{-1}|^2)[(\delta a_{-1})^2+(\delta b_1)^2]+\\\\
&(\tilde{\alpha}_0+3\beta a_0^2)(\delta a_0)^2+
\beta Re[\bar{\psi}_1^2(\delta\psi_1)^2
+\bar{\psi}_{-1}^2(\delta\psi_{-1})^2]+\\\\
&-\gamma_1(\delta a_1+\delta a_{-1})\delta a_0
-\gamma_2(\delta a_1\delta a_{-1}+(\delta b_1)^2)
.
\end{array}
\label{ener}
\end{equation}
Introducing variables $a_+$ and $a_-$ and their variations, we obtain
for the normal state (all fields equal zero)
\begin{equation}
\begin{array}{ll}
\delta^2{\cal{E}}=
&\frac 12(\tilde{\alpha}_1-\frac{\gamma_2}{2})(\delta a_+)^2+
\frac 12(\tilde{\alpha}_1+\frac{\gamma_2}{2})(\delta a_-)^2+\\\\
&2(\tilde{\alpha}_1+\gamma_2)(\delta b_1)^2+
\tilde{\alpha}_0(\delta a_0)^2
-\gamma_1\delta a_+\delta a_0.
\end{array}
\label{dener}
\end{equation}
The necessary and sufficient condition for stability of the normal
state is the positive definiteness of the second variation of
energy, which implies the conjuction of the following inequalities
\equln{\tilde{\alpha}_0>0
,}{N0}
\equln{\tilde{\alpha}_0>\frac 12(\delta-
|\gamma_2|)
,}{N1}
\equln{\tilde{\alpha}_0>\alpha_{02}
,}{N2}
where $\alpha_{02}$ is defined by the eqn. (\ref{al2}).\\
Let us introduce the material parameters plane with coordinates
$\gamma_2$ and $\delta$. For every point of the plane
($\gamma_2,\delta$) there
exists a stable normal state, depending on the temperature.
The highest temperature in which (for a given material) the normal
state becomes unstable determines the onset of superconductivity.
Instability of the normal state implies stability of a
superconducting state.
\section{Stability of the superconducting states}
We shall say that a solution from a given class has
the property of internal stability if it is stable with respect
to variations preserving the class.\\

For the classes with real solutions we have
\begin{equation}
\begin{array}{ll}
\delta^2{\cal{E}}=
&(\tilde{\alpha}_1+3\beta a_1^2)(\delta a_1)^2+
(\tilde{\alpha}_1+3\beta a_{-1}^2)(\delta a_{-1})^2+\\\\
&(\tilde{\alpha}_0+3\beta a_0^2)(\delta a_0)^2+
[\tilde{\alpha}_1+\beta(a_1^2+a_{-1}^2)-\gamma_2](\delta (b_1)^2+\\\\
&-\gamma_1(\delta a_1+\delta a_{-1})\delta a_0
-\gamma_2\delta a_1\delta a_{-1}.
\end{array}
\label{enAC}
\end{equation}
hence for the class (A), taking into account the solution (\ref{a1})
\equln{\delta^2{\cal{E}}=-(\tilde{\alpha}_1+\gamma_2)(\delta a_+)^2
-(\tilde{\alpha}_1+\frac{\gamma_2}{2})(\delta a_-)^2
+\tilde{\alpha}_0(\delta a_0)^2-\gamma_1\delta a_+\delta a_0
.}{d2eA}
Note that the coefficient by $(\delta b_1)^2$ equals zero.
The conditions for stability of mode (A) read
\equln{\tilde{\alpha}_0<\frac{\delta}{2}-\gamma_2
,}{stA1}
\equln{\tilde{\alpha}_0<\frac 12(\delta+\gamma_2)
,}{stA2}
\equln{\tilde{\alpha}_0(\tilde{\alpha}_0-(\frac{\delta}{2}-\gamma_2))
+\frac 14\gamma_1^2<0
,}{stA5}
and the condition (\ref{N0}).
To fulfill (\ref{stA1}) and (\ref{stA5}) the following restriction
of the material parameters is necessary
\equln{\frac{\delta}{2}-\gamma_2>|\gamma_1|
.}{stA3}
For such materials, the real roots of the polynomial from (\ref{stA5})
have the form
\equln{\alpha_{1,2}=\frac 12(\frac{\delta}{2}-\gamma_2\pm
\sqrt{(\frac{\delta}{2}-\gamma_2)^2-\gamma_1^2}\,)
.}{stA4}

   Within the class (C1) the second variation of energy has the form
\begin{equation}
\begin{array}{ll}
\delta^2{\cal{E}}=
&(2\tilde{\alpha}_1+\gamma_2+3\beta a_+^2)(\delta a_-)^2+
(2\tilde{\alpha}_1-\gamma_2+6\beta a_+^2)(\delta a_+)^2+\\\\
&(2\tilde{\alpha}_1-\gamma_2+2\beta a_+^2)(\delta b_1)^2+
(\tilde{\alpha}_0+3\beta a_0^2)(\delta a_0)^2-
2\gamma_1\delta a_+\delta a_0.
\end{array}
\label{enC1}
\end{equation}
and the stability conditions read
\equln{2\tilde{\alpha_0}-\delta+\gamma_2+3\beta a_+^2>0
,}{stC1a}
\equln{2\tilde{\alpha_0}-\delta-\gamma_2+2\beta a_+^2>0
,}{stC1b}
\equln{\tilde{\alpha_0}+3\beta a_0^2>0
,}{stC1c}
\equln{(2\tilde{\alpha_0}-\delta-\gamma_2+6\beta a_+^2)
(\tilde{\alpha_0}+3\beta a_0^2)-\gamma_1^2>0
,}{stC1d}

   The conditions of positive definiteness of the second variations  
for the class (C2)
\begin{equation}
\begin{array}{ll}
\delta^2{\cal{E}}=
&(2\tilde{\alpha}_1+\gamma_2+3\beta(a_+^2+2a_-^2))(\delta a_-)^2+
(2\tilde{\alpha}_1-\gamma_2+3\beta(2a_+^2+a_-^2))(\delta a_+)^2+\\\\
&(2\tilde{\alpha}_1-\gamma_2+2\beta(a_+^2+a_-^2))(\delta b_1)^2+
(\tilde{\alpha}_0+3\beta a_0^2)(\delta a_0)^2-\\\\
&2\gamma_1\delta a_+\delta a_0+6\beta a_+a_-\delta a_+\delta a_-,
\end{array}
\label{enC2}
\end{equation}
and for the class (B) 
\begin{equation}
\begin{array}{ll}
\delta^2{\cal{E}}=
&(2\tilde{\alpha}_1+\gamma_2+\beta(3a_+^2+2b_1^2))(\delta a_-)^2+
(2\tilde{\alpha}_1-\gamma_2+2\beta(3a_+^2+b_1^2))(\delta a_+)^2+\\\\
&(2\tilde{\alpha}_1-\gamma_2+2\beta(a_+^2+3b_1^2))(\delta b_1)^2+
(\tilde{\alpha}_0+3\beta a_0^2)(\delta a_0)^2-\\\\
&2\gamma_1\delta a_+\delta a_0+2\beta a_+b_1\delta a_+\delta b_1,
\end{array}
\label{enB}
\end{equation}
are necesarry and sufficient to ensure the stability in the
corresponding classes. Those modes, however, do not appear at
the onset of the superconductivity in the system under
consideration.
\section{Conclusion}
   1. The isolated atomic planes are all described by the model
2D GL with the same parameters $\alpha_0$ and $\beta$.
The parameter $\alpha_0$ depends on the temperature. 
If $\alpha_0>0$, then the 2D state of such a plane is N (normal),
in the oposite case it is the state S (superconducting). We
interpret the parameter $\alpha_0$ as a mesure of empiric
temperature, and introduce the notation $\tau=\tau_0+\alpha_0$.
The result for an isolated plane: state N for
$\tau>\tau_0$, state S for $\tau<\tau_0$.\\

   2. Each material is charcterised by two empiric temperatures
$\tau_{A}$ and $\tau_{C}$ defined by the formulae
\equln{\tau_{A}=\tau_0-\zeta_1+\frac 12(\delta-\gamma_2)
,}{tAN}
and
\equln{\tau_{C}=\tau_0-\zeta_1+\alpha_{02}
,}{tCN}
where $\alpha_{02}$ is given by the eqn. (\ref{al2}).
As long as $\tau>max(\tau_A, \tau_C)$, the normal state is stable.
Below this limit a stable superconducting mode apeares.
If  $\tau_{A}>\tau_{C}$, the mode is A; in the
oposite case - it is the mode C. The other superconducting modes
do not appear in the onset.\\

   3. On the material plane ($\gamma_2,\delta$)
the limiting curve between the regions of onset A and onset C is placed
at the half-plane 
$\gamma_2<0$, and given by the formula
\equln{\delta=\gamma_2-\frac{\gamma_1^2}{\gamma_2}
.}{llim}

   4. The solutions belonging to the class (A) exist in the whole
plane $(\gamma_2,\delta)$. The unique condition for their existence
is $\tilde{\alpha}_0<\frac 12 (\delta-\gamma_2$), hence
$\tau<\tau_{A}$. This condition ensures also the internal stability
of the solutions.\\

   5. Under the above circumstances the necessary and sufficient
condition for the stability of solutions from the class (A) is their
C-stability expressed by the eqn. (\ref{stA5}).
However C-stability of the solutions from the class (A) is possible
only for the materials fulfilling the inequality (\ref{stA3}).\\

   6. Then the necessary condition of such stability is
$\tilde{\alpha}_0>\alpha_1$, (with $\alpha_1$ given by
(\ref{stA4})), hence $\tau>\tau_{CA}$. Violation of the condition
implies the loss of stability on behalf of the mode (C2).\\

   7. If material is of the type (A), the solutions from the class (A)
are stable in the interval $\alpha_1<\tilde{\alpha}_0<\alpha_{02}$,
hence in the interval $\tau_{CA}<\tau>\tau_{A}$.\\

   8. If the material is of the type (C), then the solutions from
the class (A) are stable in the interval
$\alpha_1<\tilde{\alpha}_0<\alpha_2$, hence 
$\tau_{CA}<\tau<\tau_{AC}$. At the both ends the stability is lost
on behalf of the mode (C2).\\

   9. Now we shall discuss the enhancement of superconductivity
in the system under consideration. We consider the enhancement from
the following two points of view:\\
\indent a) The enhancement due to long distance coupling with respect
to the\\
\indent\indent short distance ones,\\
\indent   b) The enhancement of 3D superconductivity with respect to
the 2D.\\

  10. Consider first increments of the onset temperatures $\tau_A$
and $\tau_C$ due to long distance couplings. Let $\tau_{A0}$ and
$\tau_{C0}$
denote the onset temperatures $\tau_{A}$ and $\tau_{C}$ for the first
grade material, i.e. for $\zeta_2=0$ and $\gamma_2=0$. Then, from the
equations
(\ref{tAN})--(\ref{tCN}) it follows that
\equln{\tau_{A}-\tau_{A0}=-\frac 12(\zeta_2+\gamma_2)
,}{t0N}
and
\equln{\tau_{C}-\tau_{C0}=-\frac 14[
\sqrt{(\zeta_1-(\zeta_2-\gamma_2))^2+8\gamma_1^2}-
\sqrt{\zeta_1^2+8\gamma_1^2}+(\zeta_2-\gamma_2)]
.}{t0C}
  11. In consequence of the above formulae, the mode A
superconductivity is enhanced provided that
\equln{\zeta_2+\gamma_2<0
.}{tneg}
Similarly, due to the fact, that the right hand side of the
eqn. (\ref{t0C}) is a monotically increasing function of the
difference $\zeta_2-\gamma_2$, the mode C supeconductivity is
enhanced provided that
\equln{\zeta_2-\gamma_2>0
.}{tpos}
In consequence, negative values of $\gamma_2$ are in favour of
both the modes A and C. At the same time, negative values of
$\zeta_2$ enhance the mode A and tend to supress the mode C
of superconductivity.\\

   12. For some materials, charactised by appropriate values of
the parameters $\zeta_1$, $\zeta_2$ and $\gamma_2$, 
the formulae (\ref{tAN}) and (\ref{tCN}) can give
the onset
temperatures $\tau_A$ and/or $\tau_C$ greater than the 2D critical
temperature $\tau_0$. In such a situation the out-of-plane
superconductivity appears in spite of the fact that all the layers
remain in overcritical in-plane states.

\section{Acknowledgements}
This work was partially supported by the Science Research
Committee (Poland) under grant No. 5 TO7A 040 22.\\\\
\thebibliography{12}
\bibliography{}
\end{document}